\documentstyle[11pt,aaspp4,flushrt]{article}

\newcommand{\partials}[2]{\frac{\partial #1}{\partial #2}}
\newcommand{\pfrac}[2]{\left(\frac{#1}{#2}\right)}
\newcommand{\afrac}[2]{\left|\frac{#1}{#2}\right|}

\newcommand{\Meszaros}{M\'esz\'aros}

\newcommand{\rt}{\tilde{r}}
\newcommand{\tp}{\tilde{t}}
\newcommand{\co}{{co}}
\newcommand{\obs}{{\rm obs}}
\newcommand{\D}{{\cal D}}
\newcommand{\ext}{{\rm ext}}
\newcommand{\rad}{{\rm rad}}
\newcommand{\Min}{{\rm min}}
\newcommand{\Max}{{\rm max}}
\newcommand{\peak}{{\rm peak}}

\newcommand{\dt}{{t_\obs}}

\newcommand{\Int}{{\displaystyle \int}}

\newcommand{\G}{{\Gamma}}
\newcommand{\g}{{\gamma}}
\newcommand{\gdot}{{\dot{\gamma}}}
\newcommand{\e}{{\epsilon}}
\newcommand{\ep}{{\tilde{\varepsilon}}}

\newcommand{\M}{{\cal M}}

\begin{document}

\title{Synchrotron and SSC Emission and the Blast-Wave Model of 
Gamma-Ray Bursts}

\author{James Chiang\altaffilmark{1} and Charles D. Dermer}
\affil{E. O. Hulburt Center for Space Research, Code 7653, Naval Research 
Laboratory, Washington DC 20375-5352}
\altaffiltext{1}{NAS/NRC Research Associate}

\begin{abstract}
We investigate the dynamics and radiation from a relativistic
blast-wave which decelerates as it sweeps up ambient matter.  The bulk
kinetic energy of the blast-wave shell is converted into internal
energy by the process of accreting external matter.  If it takes the
form of non-thermal electrons and magnetic fields, then this internal
energy will be emitted as synchrotron and synchrotron self-Compton
radiation.  We perform analytic and numerical calculations for the
deceleration and radiative processes and present time-resolved spectra
throughout the evolution of the blast-wave.  We also examine the
dependence of the burst spectra and light curves on various parameters
describing the magnetic field and non-thermal electron distributions.
We find that for bursts such as GRB~910503, GRB~910601 and GRB~910814,
the spectral shapes of the prompt gamma-ray emission at the peaks in
$\nu F_\nu$ strongly constrain the magnetic fields in these bursts to
be well below ($\la 10^{-2}$) the equipartition values.  These
calculations are also considered in the context of the afterglow
emission from the recently detected gamma-ray burst counterparts.
\end{abstract}

\section{Introduction}

The recent Beppo-SAX observations of fading X-ray afterglows
coincident with the positions of gamma-ray bursts GRB~970228 (Costa et
al.\ 1997\markcite{Costa97}) and GRB~970508 (Piro et al.\
1998\markcite{Piro98}) have led to the first identifications of
possible burst counterparts in the optical wave band (van Paradijs et
al.\ 1997\markcite{vanParadijs97}; Djorgovski et al.\
1997\markcite{Djorgovski97}).  In the optical spectrum associated with
the latter burst, absorption lines have been detected with a redshift
of $z = 0.835$ providing, for the first time, direct evidence for a
gamma-ray burst at cosmological distances (Metzger et al.\
1997\markcite{Metzger97}). The gamma-ray fluence of GRB~970508
measured by BATSE is $\sim 3 \times 10^{-6}\,$erg~cm$^{-2}$
(Kouveliotou et al.\ 1997\markcite{Kouveliotou97}).  Therefore, if the
absorption line measurements give a lower limit on the redshift of the
burst, this implies an isotropic burst energy of $>
10^{51}\,$erg~s$^{-1}$.  An impulsive event releasing this amount of
energy in a compact region naturally leads to a fireball and thence to
a relativistic blast-wave.

Blast-wave models for gamma-ray bursts have been examined previously
in the literature in several contexts.  The most extensive body of
work on this topic has been produced by \Meszaros, Rees and
collaborators (e.g., \Meszaros\ \& Rees 1992a; Rees \& \Meszaros\ 1992;
\Meszaros, Laguna \& Rees 1993; Wijers, Rees \& \Meszaros\ 1997;
Panaitescu \& \Meszaros\ 1998a, 1998b; et al.).  Other recent papers
include Sari \& Piran (1995), Vietri (1997), Waxman (1997), and Katz
\& Piran (1997).  The basic fireball/blast-wave model consists of some
triggering event---either coalescing neutron stars or black holes
(\Meszaros\ \& Rees 1992b) or the collapse of a massive star
(Paczy\'nski 1998) or a failed type II supernova (Woosley
1993)---depositing a large amount of energy, $E_0 \sim
10^{51}$--$10^{55}$~ergs, in a small region with radius $r_0 \sim
10^6$--$10^7$~cm.  Because of the unavoidable presence of baryonic
material, it is expected that the initial fireball energy will be
transformed into kinetic energy of these baryons rather than escape as
radiation.  This material expands until the internal motions of the
baryons become sub-relativistic in the co-moving frame of the
material, at which point it forms a cold shell with bulk Lorentz
factor $\G_0 \simeq E_0/M_0 c^2$, where $M_0$ is the rest mass of the
contaminating baryons.

This shell continues to expand freely into the surrounding ambient
medium until the integrated momentum impulse upon the shell by the
swept-up matter is equal to the rest mass of the original material,
$\G_0 4\pi r_d^3 \rho_\ext /3 \approx M_0$ where $\rho_\ext$ is the
mass density of the external medium.  This defines the so-called
deceleration radius $r_d$ (Rees \& \Meszaros\ 1992).  Beyond this
radius, the shell can no longer be regarded as freely expanding, and
the bulk kinetic energy of the blast-wave begins to be reconverted
into internal energy.  If this internal energy is radiated promptly,
then the deceleration of the blast-wave shell can be approximately
described by $\G(r) \propto r^{-3}$, and the expansion is said to be
in the radiative regime.  On the other hand, if the internal energy is
radiated on a time scale which is long compared to the expansion time
scale, then the expansion is in the non-radiative regime, and $\G(r)
\propto r^{-3/2}$.  In either case, for large initial Lorentz factors,
$\G_0 \sim 10^2$--$10^3$, relativistic effects (e.g., Rees 1966)
compress the time scale for the radiation such that the bulk of the
blast-wave energy is emitted in the first tens of seconds in the
observer's frame following the initial detonation event, thus
producing the observed gamma-ray burst.

Several authors have pointed out that well after the prompt gamma-ray
burst event the blast-wave shell will continue to decelerate and
radiate (Vietri 1997; Waxman 1997).  The recent detections by the
X-ray, optical and radio communities of the aforementioned fading
X-ray, optical and radio counterparts within the error boxes of GRBs
appear to support this picture.  Furthermore, model estimates of the
temporal decay of these transients yield time-dependencies which agree
with those observed, $F_\nu \sim t^{-1}$ (Wijers et al.\ 1997); and
for the one of the bursts for which optical data are available
(GRB~970508), the optical spectral index is consistent with
synchrotron emission from a power-law distribution of electrons,
$dN/d\g \propto \g^{-p}$ with $p \simeq 2$--2.3 (Djorgovski et al.\
1997) indicating, for example, a shock-accelerated electron
population.

Despite the successes of the blast-wave model in accounting for the
prompt burst properties and its prediction of fading afterglows,
several important theoretical questions must be addressed in order to
have a reasonably complete model:
\begin{itemize}
\item What is the nature of the coupling between the electrons, protons
      and magnetic field (Panai\-tescu \& \Meszaros\ 1998b)?  To what
      extent can these components be in equipartition given that only
      the electrons can efficiently radiate away their energy?
\item How does the magnetic field change as the blast-wave decelerates?  
      If it is initially formed through equipartition processes, but
      is not strongly coupled to the non-thermal electron energy
      density, how does it evolve?  
\item What is the nature of the acceleration mechanism?  Are the 
      electrons energized by repeated diffusion across the shocks
      themselves or by gyroresonant scattering with disturbances in
      the post-shock turbulent MHD fluid?
\item What is the proper form for the injected electron energy
      distributions?  Is it well described by a typical power-law
      spectrum?  If so, what determines the characteristic energies of
      the particles?
\end{itemize}

In order to begin to address these questions, it is worthwhile to go
beyond the simple, though useful, back-of-the-envelope estimates which
have generally prevailed in the literature thus far.  In this paper,
we present a detailed calculation of the blast-wave deceleration and
the evolution of the magnetic fields and electron distributions under
various assumptions relevant to the above issues.  We compute model
light curves and spectra and use the available afterglow data to help
discriminate between the various options.  The format of the paper is
as follows: In \S~2, we describe the basic dynamics of an impulsively
driven blast-wave which decelerates by accretion of ambient material.
In \S~3, we discuss the physical processes responsible for producing
the observed radiation including prescriptions for magnetic field
generation, the formation of non-thermal particle distributions and
the relevant radiation processes.  The numerical procedures for
computing the deceleration of the blast-wave and the integration of
the blast-wave shell emission are described in \S~4.  Model spectra
and light curves for gamma-ray bursts are presented and analyzed in
\S~5.  Lastly, in \S~6, we discuss these results, explore prospects
for further research and present our conclusions.

\section{Deceleration of Impulsively Driven Blast-Waves}

We assume that a shell of material, with initial proper mass $M_0$, is
injected with a bulk Lorentz factor $\G_0$ and that its subsequent
deceleration is determined by energy and momentum conservation as it
accretes mass in the form of matter and internal energy from the
ambient medium and loses inertia via radiative processes such as
synchrotron or inverse Compton emission.  The internal energy of the
material is assumed to be initially negligible compared to its rest
mass energy.  This corresponds to a time well after the random
internal energy of the initial fireball has been converted into the
bulk kinetic energy of the spherically expanding shell.  We model the
blast-wave as having a cross-sectional area which is a function of the
distance traveled $r$.  This functional dependence can be parametrized
as a power-law: $A(r) = A_0 (r/r_0)^a = 4 \pi f_b r^a$ where $r_0$ and
$A_0$ are the initial radius and surface area of the blast-wave,
respectively, and $f_b$ is a collimation factor to account for
possible asymmetry of the explosion.  In the calculations presented
throughout this paper, we consider a spherical blast-wave with $A(r) =
4\pi r^2$.

A straight-forward application of energy and momentum conservation
gives the equation of motion of the blast-wave shell as it sweeps up
ambient matter.  We obtain the following expression (Blandford \&
McKee 1976):
\begin{equation}
\frac{d\G}{dm} = -\frac{\G^2-1}{M},
\label{equation_of_motion}
\end{equation}
where $\G = \G(r)$ is the bulk Lorentz factor of the material, $M =
M(r)$ is the total mass including internal kinetic energy, $dm(r) =
A(r)\rho(r) dr$ is the {\em rest} mass swept-up in a distance $dr$,
and the distance traveled $r$ is measured in the stationary frame of
the explosion.  This frame will be referred to hereafter as the lab
frame.

\subsection{Radiative and Non-radiative Limits}

As the blast-wave shell decelerates, its bulk kinetic energy is
converted to internal kinetic energy.  If this internal energy is not
radiated away then it makes a significant contribution to the inertia
of the blast-wave shell.  In this case, the mass $M$, in
equation~\ref{equation_of_motion}, is given by
\begin{equation}
M_{nr}(r) = M_0 + \int_{r_0}^r \G(\rt) A(\rt) \rho(\rt) d\rt,
\label{non-radiative_mass}
\end{equation}
where the subscript $nr$ identifies this as the non-radiative case.
In the radiative case, the internal kinetic energy is immediately
lost, and the inertia of the blast-wave shell consists only of the
rest mass:
\begin{equation}
M_r(r) = M_0 + \int_{r_0}^r A(\rt) \rho(\rt) d\rt.
\label{radiative_mass}
\end{equation}
Note that since any radiation is assumed to be isotropic in the
co-moving frame of the blast-wave shell, the bulk Lorentz factor is
unaffected by this radiation and equation~\ref{equation_of_motion}
still applies.

In the radiative case, $dM_r = dm$, the equation of motion 
(eq.~\ref{equation_of_motion}) is separable, and the resulting 
integrals are elementary.  The solution is 
\begin{equation}
\G(r) = \frac{(M_r(r)/M_0)^2(\G_0 + 1) + \G_0 - 1}
             {(M_r(r)/M_0)^2(\G_0 + 1) - \G_0 + 1}
\label{radiative_soln}
\end{equation}
(Blandford \&~McKee 1976\markcite{Blandford76}).  In the non-radiative
case, the deceleration is described by a pair of coupled ordinary
differential equations:
\begin{eqnarray}
\frac{d\G}{dr}     &=& -\frac{A(r)\rho(r)(\G(r)^2 - 1)}{M_{nr}(r)},\\
\frac{dM_{nr}}{dr} &=& \G(r)A(r)\rho(r).
\label{dMnrdr}
\end{eqnarray}
These equations also have an analytic solution.  The general expression
for the mass is
\begin{equation}
M_{nr}(r) = \left[M_0^2 + 2 \G_0 M_0 M_{sw}(r) + M_{sw}(r)^2\right]^{1/2},
\end{equation}
where $M_{sw}(r) = \int_{r_0}^r d\rt A(\rt) \rho(\rt)$ is the swept-up
{\em rest} mass of the ambient material.  The bulk Lorentz factor
is given by 
\begin{equation}
\G(r) = \frac{M_{sw}(r) + \G_0 M_0}{M_{nr}(r)}.
\end{equation}
For a spherical blast-wave in a uniform external medium, $A(r)\rho(r)
= 4\pi r^2 \rho_0$.  For $r \gg r_0$, the mass of the blast-wave shell
is given by
\begin{equation}
M_{nr}(r) = \left[M_0^2 + 2 \G_0 M_0 \frac{4\pi}{3}r^3\rho_0
            + \left(\frac{4\pi}{3}r^3\rho_0\right)^2\right]^{1/2};
\label{nonradiative_mass}
\end{equation}
and the bulk Lorentz factor is
\begin{equation}
\G(r) = \frac{1}{M_{nr}(r)}\left[\frac{4\pi}{3}r^3\rho_0 + \G_0 M_0\right].
\end{equation}

From the above expression for the mass in the non-radiative case, we
identify three regimes for the evolution of the blast-wave depending
on which of the three terms dominates in
equation~\ref{nonradiative_mass}.  For $M_0 > 2 \G_0 M_{sw}$,
\begin{eqnarray}
M_{nr}(r) & \approx & M_0,\\
\G(r)     & \approx & \G_0.
\end{eqnarray}
This regime corresponds to the initial period of free-expansion.
For $2 \G_0 M_{sw} > M_0 > M_{sw}/2\G_0$,
\begin{eqnarray}
M_{nr}(r) &\approx & \left[2 \G_0 M_0 M_{sw}\right]^{1/2},\\
\label{Mnr_mass}
\G(r)     &\approx & \G_0\left[\left(\frac{M_0}{2\G_0 M_{sw}}\right)^{1/2}
                     + \frac{1}{\G_0}\left(\frac{M_{sw}}
                       {2\G_0 M_0}\right)^{1/2}\right]\\
          &\propto & r^{-3/2}.
\end{eqnarray}
Here we recover the familiar power-law $r^{-3/2}$ dependence of the
bulk Lorentz factor which is often quoted for a non-radiative
relativistic blast-wave decelerating in a uniform medium.  Finally,
when $M_{sw} > 2\G_0 M_0$ the bulk motion of the blast-wave is
non-relativistic, and we have
\begin{eqnarray}
M_{nr}(r) & \approx & M_{sw}(r),\\
\G(r)     & \approx & 1 + \G_0\frac{\M_0}{M_{sw}(r)}.
\end{eqnarray}
The velocity of the blast-wave inferred in this limit, using the
approximation $\G \simeq 1 + \beta^2/2$, is $\beta =
\sqrt{\G_0 M_0/M_{sw}} \propto r^{-3/2}$, which agrees with the
adiabatic solution for non-relativistic blast-waves (Lozinskaya 1992).

Energy conservation gives the rate of accreted kinetic energy as a
function of time in the lab frame:
\begin{equation}
\frac{dE}{dt} = c^3 A(r)\rho(r)\beta (\G^2 - \G).
\end{equation}
This expression follows directly from the equation of motion
(eq.~\ref{equation_of_motion}) and therefore applies regardless of
whether the blast-wave is radiative or non-radiative.  Since the above
quantity is a Lorentz invariant, it also gives the injected power in
the co-moving frame of the blast-wave (Blandford \& McKee
1976\markcite{Blandford76}).

\subsection{Partially Radiative Regimes}
In general, the evolution of the blast-wave shell will be intermediate
between the radiative and non-radiative limits.  If we denote the
radiated power in the co-moving frame as $dE_{\rm rad}/dt$, then the
relativistic mass of the material is
\begin{equation}
M(r) = M_0 + \int_{r_0}^r d\rt \G(\rt)\rho(\rt)A(\rt) 
       - \int_{0}^t d\tp\frac{1}{c^2}\frac{dE_\rad}{d\tp},
\end{equation}
where the latter integral is performed over co-moving time and
represents the relativistic mass lost to radiation.  Converting to the
lab frame coordinate using $d\rt = c\beta\G d\tp$, and differentiating
wrt $r$, we obtain a more general form for equation~\ref{dMnrdr}:
\begin{equation}
\frac{dM}{dr} = \G(r)\rho(r)A(r) - \frac{1}{c^3\beta(r)\G(r)}
                \frac{dE_\rad}{dt}.
\label{dMdr}
\end{equation}
This expression must be integrated numerically in conjunction with the
equation of motion (eq.~\ref{equation_of_motion}). The radiated power
$dE_{\rm rad}/dt$ is determined by the co-moving particle
distributions and the relevant energy loss mechanisms.  We address
these issues in the following sections.

\section{Physical Processes}

\subsection{Particle Injection Distributions}
We assume that some fraction, $\xi_e$, of the accreted kinetic energy
is injected directly into non-thermal electrons.  Energization of
these electrons may occur via acceleration by shocks (see e.g.,
Ellison, Baring \& Jones 1995) or by gyroresonant scattering with
plasma-wave turbulence induced by the swept-up material (Dermer,
Miller \& Li 1996 \markcite{Dermer96}).  However, since we do not
model these processes in detail, we adopt a power-law injection
spectrum for the electron momenta, $dN/dp \sim p^{-s}$.  In terms of
the initial electron Lorentz factors, $\g_i$, this injection spectrum
per unit co-moving time is
\begin{equation}
\frac{dN}{d\g dt_i} = \frac{dN_0}{dt_i} \g_i(\g_i^2 - 1)^{(s+1)/2}.
\label{injection_function}
\end{equation}
Here the time-dependent normalization of the injection spectrum is
given by the prescribed fraction of the power available as kinetic
energy:
\begin{equation}
\frac{dN_0}{dt_i} = \frac{\xi_e dE/dt}
                    {m_e c^2 \Int_{\!\!\!\g_{i,\Min}}^{\g_{i,\Max}}
                     (\g_i^2 - \g_i)(\g_i - 1)^{(s+1)/2}d\g_i}.
\end{equation}
The minimum injected electron Lorentz factor in the co-moving frame
should be at least as large as the bulk Lorentz factor of the
blast-wave, $\g_{i,\Min} = \G$.  However, if the electrons are in
equipartition with the protons, then the lower cut-off for the
electron distribution will be approximately the typical energy of the
incident protons in the co-moving frame, $\g_{i,\Min} = (m_p/m_e)\G$
(Katz \&~Piran 1997\markcite{Katz97}; Panaitescu \& \Meszaros\ 1998a).
We parametrize the uncertainty between these two extremes with a
parameter $\eta$:
\begin{equation}
\g_{i,\Min} = \eta \frac{m_p}{m_e}\G(r_i).
\label{gmin_equation}
\end{equation}

The upper cut-off of the electron distribution is given by balancing
the radiative losses with the power gained due to the acceleration.
By balancing acceleration time scales for both shock and gyroresonant
processes with synchrotron loss time scales, de~Jager et al.\
(1996\markcite{deJager96}; see also de~Jager
\&~Harding 1992\markcite{DeJager92}) argue that for a given
magnetic field the maximum electron Lorentz factor is:
\begin{equation}
\g_{i,\Max} \simeq 4 \times 10^7 (B/1 {\rm ~G})^{-1/2}.
\end{equation}

\subsection{Magnetic Field Prescriptions}
At a minimum, the post-shock magnetic field will be that of the
ambient external field, amplified by the compression ratio ($\sim 4$
for a strong shock) and the Lorentz transformation to the co-moving
frame: $B \sim 4\G B_{\rm pre} \sim 10^{-5} \G$~Gauss.  Here we have
assumed an ambient magnetic field intensity typical for the ISM of a
few $\mu$G.  Even for bulk Lorentz factors of $\G \sim 10^3$, this
field strength is too low for the electrons to be radiatively
efficient.  Another source of the magnetic field energy is the remnant
field which was created during the initial fireball event and which
gets carried along with the expanding blast-wave material.  However,
if flux freezing holds, i.e., $r^2B \sim$~constant, the field energy
will decrease in proportion with the internal energy of the
contaminating baryons, and by assumption we only consider the shell
after it has expanded to the point at which the shell material is cold
in the co-moving frame, hence the remnant field energy density should
also be negligible.

If gyroresonant processes mediate the approach to equipartition
between the protons and electrons, then it follows that the magnetic
field will also be in equipartition with the particles.  In true
equipartition, the bulk of the internal energy will be deposited in
components which are themselves not efficient radiators, namely the
protons and the magnetic field.  However, if strong coupling exists
between the electrons, protons and the magnetic field such that these
components remain in equipartition even as the electrons efficiently
radiate away their kinetic energy, then we would expect the dynamics
of the blast-wave to be in the radiative regime (Panaitescu \&
\Meszaros\ 1998b).  In practice, this would correspond to $\xi_e
\approx 1$ since the electrons would then be able to tap all of the
available internal energy of the blast-wave shell.  If the coupling
between the magnetic field and particles is weak, then we would expect
the internal energy in the protons and magnetic field to remain in the
blast-wave and the blast-wave dynamics should be intermediate between
radiative and non-radiative.  In either case, we would expect that the
generated magnetic field energy density will be proportional to the
energy density of the swept-up material.  This implies a magnetic
field of the form:
\begin{equation}
B = \left(32\pi\xi_B \rho_\ext\right)^{1/2} \G c,
\label{magnetic_field}
\end{equation}
where we have assumed a compression ratio of 4.  The parameter $\xi_B$
contains all the uncertainties regarding particle-field coupling and
the degree to which equipartition is achieved and maintained.  In
principle, $\xi_B$ will change with time depending on the complicated
plasma physics processes taking place within the blast-wave.  However,
we will treat it as a constant and address the implications of this
simplification in our discussion below.

\subsection{Radiation Processes}
The radiation processes which we consider are synchrotron and
synchrotron self-Compton.  An expression for optically thin emission
spectra for these processes are given by Dermer, Sturner \&
Schlickeiser (1997)\markcite{Dermer97} where they use a
$\delta$-function approximation for the synchrotron spectrum of a
single electron which is centered at the cyclotron frequency and which
is normalized by the total synchrotron power.  When convolved with a
smooth distribution of electron energies, this approximation provides
a reasonably accurate representation of the optically thin synchrotron
emissivity:
\begin{eqnarray}
j_{\rm syn}(\e) &=& \frac{dE}{dt dV d\e d\Omega}\nonumber\\
                &\simeq& \frac{c\sigma_T u_B}{6\pi\e_B} \pfrac{\e}{\e_B}^{1/2}
                    \frac{dN}{dVd\g}\left[\pfrac{\e}{\e_B}^{1/2}\right].
\label{synch_spectrum}
\end{eqnarray}
Here $\e = h\nu/m_e c^2$, $\e_B = B/B_{\rm crit} = B/4.414\times
10^{13}$~G, $\sigma_T$ is the Thomson cross-section, and $dN/dVd\g$
is the electron distribution function.  Unfortunately, this
approximation breaks down at the end-points of the electron
distribution.  In order to obtain a more accurate photon spectrum, we
apply a more sophisticated calculation (Crusius \&~Schlickeiser 1986;
1988) appropriate for an isotropic distribution of electrons.
However, we note that since the sharp electron cut-offs are themselves
approximations, the spectra and light curves due to electrons at
these cut-offs should be regarded with caution.

Using a similar $\delta$-function approximation, Dermer et al.\ (1997)
also derive an expression for the first-order SSC emissivity in the
Thomson regime:
\begin{equation}
j_{\rm ssc}(\e) \simeq \frac{c\sigma_T^2 u_B r_b \e^{1/2}}{9\pi\e_B^{3/2}}
                  \int_{\e_B}^{\Min(\e,1/\e)} d\ep \ep^{-1}
                  \frac{dN}{dVd\g}\left[\pfrac{\ep}{\e_B}^{1/2}\right]
                  \frac{dN}{dVd\g}\left[\pfrac{\e}{\ep}^{1/2}\right].
\label{ssc_spectrum}
\end{equation}
The factor $r_b$ comes from the characteristic escape time ($\sim
r_b/c$) for synchrotron photons to leave a homogeneous spherical blob.
In the case of a blast-wave shell we take $r_b$ to be the thickness of
the shell in the co-moving frame.

\section{Numerical Procedures}

\subsection{Evolution of the Non-Thermal Particle Distributions}

The non-thermal particle distribution is governed by the 
equation of continuity in energy space:
\begin{equation}
\partials{}{t}\pfrac{dN}{d\g} 
   + \partials{}{\g}\left(\gdot\frac{dN}{d\g}\right)
   = \frac{dN}{dt_i d\g_i}.
\label{electron_continuity_eqn}
\end{equation}
The losses consist of radiative losses due to synchrotron and
synchrotron self-Compton emission and adiabatic losses:
\begin{eqnarray}
\gdot &=& \gdot_\rad + \gdot_{\rm adi}\\
      &=& -\frac{1}{m_e c^2}\frac{4}{3} \beta^2\g^2 c \sigma_T 
         (u_B + u_{\rm syn}(\g)) - \frac{\g}{3}\frac{\dot{V}}{V}.
\label{particle_losses}
\end{eqnarray}
The first term on the rhs corresponds to the radiative losses where
$u_{\rm syn}$ is the energy density of the synchrotron photons which
can be up-scattered by an electron with Lorentz factor $\g$ in the
Thomson limit:
\begin{equation}
u_{\rm syn}(\g) = \int_0^{1/\g} m_e c^2 d\e \e \frac{dn_{\rm syn}}{d\e dt}.
\end{equation}
Here $dn_{\rm syn}/d\e dt$ is the synchrotron photon number spectral
density.  The second term on the rhs of equation~\ref{particle_losses}
is the expansion losses where $V$ is the volume of the emitting
material.  However, since the relevant radiative loss time scales are
much shorter than the expansion time scales, we neglect the adiabatic
loss term.  We also find that bremsstrahlung losses can be neglected
compared to the synchrotron and SSC losses.

The evolution of the particle distribution is computed by integrating
equation~\ref{electron_continuity_eqn} using an implicit finite
differencing scheme (Press et al.\ 1992):
\begin{equation}
\frac{N_{i+1\,j} - N_{ij}}{t_{i+1} - t_i} +
      \frac{\gdot_{i+1\,j+1}N_{i+1\,j+1} - \gdot_{i+1\,j}N_{i+1\,j}}
           {\g_{j+1} - \g_j} = Q_{ij}.
\label{finite_diff}
\end{equation}
Here $N_{ij} = dN/d\g(t_i,\g_j)$, $\gdot_{ij} = \gdot(t_i,\g_j)$, and
the source term, $Q_{ij}$, is the injection function given by
equation~\ref{injection_function}. This implicit finite differencing
scheme yields a tridiagonal system of equations which is readily
inverted to give the electron distribution as a function of time.  We
note that the losses associated with the SSC term depend on the
electron distribution function through $u_{\rm syn}$ and must be
determined self-consistently.  This is achieved by iteration until the
electron distribution has converged.

\subsection{Self-Consistent Dynamics}

If the kinetic energy deposited in the electrons is immediately
radiated, then equation~\ref{dMdr} is equivalent to
\begin{equation}
\frac{dM}{dr} = \rho(r)A(r)\left[(\G(r) - 1)(1 - \xi_e) + 1\right].
\label{analytic_Gamma}
\end{equation}
In general, however, $dE_\rad/dt$ depends on all previous history of
the accretion process through $dN/d\g$ and the integration of the
equation of continuity (eq.~\ref{electron_continuity_eqn}), hence
$dE_\rad/dt$ contains an implicit integral over $r$.  We integrate
equation~\ref{dMdr} by taking time steps equal to the shortest energy
loss time scale $\Delta t \sim (\g/\gdot)_\Min$, and for each of these
steps, estimate $dE_\rad/dt$ using
\begin{equation}
\pfrac{dE_\rad}{dt}_{t_i} = \frac{m_e c^2}{(\Delta t)_i} 
       \int_1^\infty d\g\,\g\left[\pfrac{dN}{d\g}_{t_{i-1}} 
                                 - \pfrac{dN}{d\g}_{t_i}\right],
\label{dEraddt_est}
\end{equation}
where $t_i = t_{i-1} + \Delta t$ is the time of the $i$th step.  The
bulk Lorentz factor and total mass are then obtained by
forward-stepping: $\G_i = \G_{i-1} + \Delta r(d\G/dr)$ and $M_i =
M_{i-1} + \Delta r(dM/dr)$, where $\Delta r = c\beta\G \Delta t$.  The
estimate of the radiated power (eq.~\ref{dEraddt_est}) from this
simple forward-stepped integration is then used as input for a
variable step-size Runge-Kutta integrator and the process is repeated
until convergence.  Only for very low magnetic fields, $\la 10^{-3}$
of the equipartition value, are several additional iterations required
in order to obtain convergence.

\subsection{Computing the Observed Spectra from the Co-Moving 
Emissivities} 

Using the electron distribution function from the integration of the
equation of continuity and the expressions for the synchrotron and SSC
emissivities, we compute the radiation in the co-moving frame of the
blast-wave shell.  Because of light travel time effects, emission from
the blast-wave shell at different lab frame times contributes to the
observed flux at a given observer time.  Hence, in the observer frame,
the flux at any given time will consist of an integration of the
emission over the course of the blast-wave expansion.  If a shell
element moves radially outward from the point of the explosion with
velocity $c\beta(r)$ and at an angle $\theta = \cos^{-1}\mu$ with
respect to the observer line-of-sight, then the time delay between
radiation from this shell element at radius $r$ and the initial
explosion event is given by
\begin{equation}
\dt = \frac{1}{c}\left[\int_0^r \frac{d\rt}{\beta(\rt)} - r\cos\theta\right].
\label{time_delay}
\end{equation}
The observed photon flux is
\begin{eqnarray}
\pfrac{dN}{dA d\e dt d\Omega}_\obs
       &=& \frac{1}{4\pi d_l^2}\int_0^\infty dr\,\int_{\rm shell} dA\,\mu\,
           \pfrac{dN}{dA d\e dt d\Omega}_\co \delta(r - \rt(\dt,\mu))
           \nonumber\\
       & & \times \afrac{d\e_\co}{d\e_\obs}\afrac{dt_\co}{dt_\obs}
                  \afrac{d\Omega_\co}{d\Omega}\afrac{dA_\co}{dA},
\end{eqnarray}
where $d_l$ is the luminosity distance to the blast-wave, the
$dA$-integral is performed over the shell surface at radius $r$, the
element of area is $dA = 2\pi r^2 d\mu$, and the factor of $\mu$
accounts for the projection of the surface area of the shell.  Note
that for convenience we have taken the redshift to be $z = 0$.  It is
straight-forward to include the appropriate number of $(1+z)$ factors
in the final expressions.  The $\rt$ in the $\delta$-function
satisfies the time delay equation~(\ref{time_delay}) for given values
of $\dt$ and $\mu$.  The transformation from the co-moving frame to
the observer frame introduces a number of Jacobian factors, all of
which are powers of the Doppler factor $\D = [\G (1 -
\beta\mu)]^{-1}$: $|{d\Omega_\co}/{d\Omega}| = \D^2$, 
$|{d\e_\obs}/{d\e_\co}| =\D$, $|d\dt/{dt_\co}| = \D^{-1}$, and 
$|{dA_\co}/{dA}| = 1$.  The area is unchanged since the velocity of 
the shell is perpendicular to its surface.  Explicitly including all 
these factors and using the fact that the emission is isotropic in the 
co-moving frame, we obtain
\begin{equation}
\pfrac{dN}{dA d\e dt d\Omega}_\obs =
         \frac{1}{4\pi d_l^2} \int_{-1}^{1} \frac{1}{8\pi}
         \left[\frac{dN}{d\e_\co dt_\co}\frac{\mu d\mu}
               {[\G(1 - \beta\mu)]^2}\right]_{r = \rt(\dt,\mu)},
\end{equation}
where $dN/d\e_\co dt_\co$ is calculated from the expressions for the
synchrotron (eq.~\ref{synch_spectrum}) and synchrotron self-Compton
(eq.~\ref{ssc_spectrum}) emissivities and the integration of the
electron equation of continuity (eqs.~\ref{electron_continuity_eqn}
\&~\ref{finite_diff}).

At a time delay $\dt$ the apparent size of the afterglow image on the
sky will be given by the maximum value of the quantity $\rho =
r\sqrt{1 - \mu^2}$.  Differentiating wrt $\mu$ and using the
expression for constant time delay (eq.~\ref{time_delay}) this maximum
occurs at $\beta = \mu$, so that $\rho_\Max = \rt/\G(\rt)$.  The
quantity $\rt$ is found by solving the equation
\begin{equation}
\rt \beta(\rt) = \int_0^{\rt} \frac{d\rt}{\beta(\rt)} - c\dt.
\end{equation}
For constant $\beta$, the solution is $\rt = c\G^2\beta\dt$ yielding
the standard result $\rho_\Max = c\G\beta\dt$.  For a blast-wave shell
in a uniform external medium whose Lorentz factor can be described by
$\G = \G_0$ for $r < r_d \equiv (3M_0/8\pi\G_0\rho_0)^{1/3}$, and $\G
= \G_0(r/r_d)^{-\zeta}$ for $r > r_d$, where $\zeta \simeq 3/2$ in the
non-radiative case and $\zeta \simeq 3$ in the radiative case, as long
as $\G \gg 1$ and $\dt > (2\zeta/(2\zeta+1))\dt_{,0}$ where $\dt_{,0} =
r_d/2\G_0^2 c$, the apparent size is
\begin{equation}
\rho_\Max \simeq \frac{r_d}{\G_0}
            \left[\frac{1}{2(\zeta+1)}\left((2\zeta+1)\frac{\dt}{\dt_{,0}} 
             - 2\zeta\right)\right]^{(2\zeta+1)/(\zeta+1)}.
\label{burst_size}
\end{equation}
In the radiative and non-radiative cases, $(2\zeta+1)/(\zeta+1) = 4/7$
and $5/8$ respectively (cf.\ Waxman, Kulkarni \& Frail 1997).  For
purposes of illustration, we take an intermediate value,
$(2\zeta+1)/(\zeta+1) = 3/5$, and obtain
\begin{equation}
\rho_\Max \simeq 1.4\times 10^{13} \pfrac{\G_0}{10^3}^{-1/15}
                      \pfrac{E_0}{10^{51}\,{\rm ergs}}^{2/15}
                      \pfrac{n_\ext}{1\,{\rm cm}^{-3}}^{-2/15}
                      \pfrac{\dt}{1\,{\rm sec}}^{3/5} \mbox{~cm}.
\label{burst_size_est}
\end{equation}
For the radio afterglow associated with GRB~970508, Frail et
al.\ (1997) find that the decline of the 8.64~GHz flux variability due
to interstellar scintillation implies a source size of $\rho_\Max
\approx 10^{17}\,$cm at $\dt \approx 1$~month.  Using
equation~\ref{burst_size_est} above, this observation is consistent
with a spherical blast-wave with initial energy $E_0 \sim
10^{51}$~ergs expanding in an external medium with density $n_\ext
\sim 1$~cm$^{-3}$.  However, the very weak dependence of
$\rho_\Max$ on $\G_0$, $E_0$, and $n_\ext$ means that any of these
parameters could vary by as much as two orders of magnitude and not
substantially change this result.  The significance of
equations~\ref{burst_size} \&~\ref{burst_size_est} is the dependence
of the apparent size with time.  For a sufficiently bright radio
afterglow, the blast-wave hypothesis could be tested by measuring the
time dependence of the amplitude of the radio scintillation to
determine if it conforms with the above prediction for the apparent
size as a function of time (Goodman 1997; Frail et al.\ 1997; Waxman
et al.\ 1997).

\section{Model Light Curves and Spectra}

We test our implementation of the above calculations by comparing our
computer program's output with the analytic results presented by
Dermer \& Chiang (1998) for the case of a radiative blast-wave and a
constant magnetic field.  For these calculations, we use $\G_0 = 300$,
$B = 1$~G, an injection index of $s = 2$, and the blast-wave is
assumed to be radiative.  Figure~\ref{test_case}a shows the spectra,
$\e^2 dN/d\e dt$, as seen by an observer at times $\dt = 1, 10, 10^2,
\dots$~seconds for the numerical (solid curves) and analytic (dashed)
calculations.  The vertical lines indicate radio (2.4~GHz, dotted),
optical (1~eV, solid), X-ray (1~keV, dashed) and gamma-ray (1~MeV,
dot-dashed) energies.  In figure~\ref{test_case}b, the light curves
corresponding to these energies are plotted, with the results from the
analytic calculation over-plotted as symbols.  The agreement is
reasonably good, and the discrepancies are due to the fact that the
analytic calculations use $\G(r) = \G_0 (r/r_d)^{-3}$ for the bulk
Lorentz factor rather than equation~\ref{radiative_soln}, somewhat
different expressions for the electron energy losses due to
synchrotron radiation, the electron distribution function and the
$\delta$-function approximation for the synchrotron emissivity.  In
addition, the use of the finite-difference method for calculating the
evolution of the electron energies results in less sharp features
(e.g., rounded high energy cut-offs) in the distributions.

In the upper panel of figure~\ref{surface_brightness}, we plot the
surface brightness in the radio waveband as a function of
perpendicular distance, $\rho$, for this calculation.  As before, a
curve is shown for time delays $\dt = 1, 10, 10^2,\ldots$~seconds
(plotted top to bottom).  Each curve is computed as a function of
observer angle $\theta = \cos^{-1}\mu$, thus the surface brightness is
double-valued and turns around at $\rho_\Max$.  The flat portion
within $\rho_\Max$ results from emission from the leading surface of
the blast-wave shell, i.e., for $\theta \le \G(\rt)$.  In the lower
panel, we plot $\rho_\Max$ for each time delay, $\dt$, along with the
curve given by equation~\ref{burst_size} using $E_0 = 10^{50}$~ergs,
$\G_0 = 300$, $n_\ext = 1$~cm$^{-3}$, and $\zeta = 3$.

An example of a more realistic calculation is shown in
figure~\ref{standard_calc}.  In figure~\ref{standard_calc}a, we plot
as the solid curve the bulk Lorentz factor of the blast-wave shell as
a function of distance from the central explosion.  The initial
Lorentz factor is $\G_0 = 300$, the fireball energy is $E_0 =
10^{52}$~ergs, and the ambient density is $n_\ext = 1$~cm$^{-3}$.
We use $\xi_B = 1$ and $\eta = 1$ for the minimum electron
energy (eq.~\ref{gmin_equation}), and the index of the electron
momentum distribution is $s = 3$.  The radius at which the
free-expansion phase ends and the deceleration phase begins is
consistent with the analytic estimate of $r_d = (3 E_0/8\pi m_p c^2
n_\ext \G_0^2) = 2\times 10^{16}$~cm.  The fraction of kinetic energy
made available for radiation by electrons is $\xi_e = 0.5$.  Since
this is less than unity, the evolution of the bulk Lorentz factor lies
between the two limiting cases with $\G(r) \propto r^{-1.9}$ for $r >
r_d$.  Figure~\ref{standard_calc}b shows the photon spectra, including
the synchrotron self-Compton component, at observed time delays $\dt =
1, 10, 10^2\ldots$~seconds after the initial fireball event.  The
blast-wave radii corresponding to these times are indicated as dotted
vertical lines in figure~\ref{standard_calc}a.

At early times, the shape of the synchrotron part of these spectra can
be inferred directly from the corresponding co-moving electron
distributions at the leading surface of the blast-wave (i.e., at $\mu
= 1$) since the bulk of the observed radiation originates from within
a small angle $\sim 1/\G$ around this region.  These distributions are
shown in figure~\ref{standard_calc}c with the earlier distributions
corresponding to the lower curves.  The shapes of these distributions
are determined by the momentum dependence of the particle energy loss
rates and the shape of the injection spectra.  The steep sections at
high energies, above the break but below the high energy cut-off, are
the power-law-injected electrons which have evolved to a cooled
distribution with $dN/d\g \sim \g^{-(s-1)}$ (Dermer \& Chiang 1998).
The break energy (e.g., $\g_{\rm break} \simeq 6\times10^5$ for the
$\dt = 1$~s curve) is the minimum injection energy corresponding to
the leading part of the blast-wave, $\g_{\rm break} \approx
\g_{i,\Min} = \eta(m_p/m_e)\G$, and the electrons with the smallest
energies ($\g_\Min \simeq 10^2$) are those which were injected at
early times with $\g_i = (m_p/m_e)\G_0$ but which have cooled due to
radiative losses.  The flat sections at energies less than $\g_\Min$
result from synchrotron cooling of continuously injected electron
distributions which have sharp low energy cut-offs.  The spectral
shape, $dN/d\g \sim \g^{-2}$, is due to the energy dependence of the
energy loss rate, $\gdot \propto -\g^{2}$.  This shape is independent
of the injection power-law index and hence will be a generic feature
of a synchrotron-cooled electron spectrum resulting from injection
spectra with sharp lower cut-off injection energies.  An analytic
expression for this part of the spectrum can be obtained by treating
the injection function as a $\delta$-function at the minimum injection
energy:
\begin{equation}
\frac{dN}{d\g_idt_i} \rightarrow \frac{dN_0}{dt_i}\, 
            \delta\!\left(\g_i - \eta\frac{m_p}{m_e}\G_i\right)
            \int_{\g_{i,\Min}}^{\g_{i,\Max}} d\tilde{\g_i}
            \tilde{\g_i}^{-s},
\end{equation}
where $\G_i = \G(r_i)$, $r_i$ is the radius at which this distribution
is injected and the integral on the rhs sets the normalization. We
find, in terms of observing time $\dt$,
\begin{equation}
\g^2\frac{dN}{d\g} = \pfrac{3\pi m_e \xi_e}{4\sigma_T m_p \xi_B \eta}
       \frac{r_d^2}{\G_0} \times \left\{\begin{array}{ll}
                                (\dt/\dt_{,0})^2 & \dt \le \dt_{,0}\\
                                \left[(2\zeta+1)(\dt/\dt_{,0})
                                 - 2\zeta\right]^{(\zeta+2)/(2\zeta+1)} & 
                                \dt > \dt_{,0}.
                                \end{array}\right.
\label{flat_spec}
\end{equation}
This expression is valid for $s > 2$.  The electron spectra implied by
this expression, including the cooled high energy power-law
distribution above the breaks at $\g_{i,\Min}$ and the estimated
roll-overs at low energies, are plotted as dashed curves in
figure~\ref{standard_calc}c.

In figure~\ref{standard_calc}b, the energy of the peak of the 
synchrotron spectrum corresponds to the break at high electron 
energies.  These quantities are related by 
\begin{eqnarray}
\e_\peak &\approx& \frac{B}{B_{\rm crit}} \g_{\rm break}^2 \G\\
              &\approx& \frac{\sqrt{32\pi\xi_B m_p c^2 n_\ext\G^2}}
                        {B_{\rm crit}}\eta^2\pfrac{m_p}{m_e}^2\G^3\\
              &\approx& 3\times 10^{-8} \xi_B^{1/2} n_\ext^{1/2} \eta^2 \G^4
\label{peak_energy}
\end{eqnarray}
(Katz \& Piran 1997).  Using this relation in conjunction with the
expressions for the synchrotron spectrum (eq.~\ref{synch_spectrum})
and the electron distributions (eq.~\ref{flat_spec}), we find the
temporal dependence of the peak of the synchrotron component:
\begin{equation}
\left(\e^2 \frac{dN}{d\e dt}\right)_\peak \simeq
           2\pi m_p c^2 n_\ext \xi_e r_d^2 \G_0^2 \times 
            \left\{\begin{array}{ll} (\dt/\dt_{,0})^2 & \dt \le \dt_{,0}\\
                 \left[(2\zeta +1)(\dt/\dt_{,0}) - 2\zeta\right]
                 ^{(2-2\zeta)/(2\zeta+1)} & \dt > \dt_{,0}. \end{array}\right.
\label{peak_luminosity}
\end{equation}
Here we see that the maximum of the light curve of the peak energy
occurs at $\dt_{,0} = r_d/2c\G_0^2$.  We also note that the above
expression is independent of the magnetic field equipartition
parameter.

With an injection index of $s$ yielding a power-law index for the
cooled distribution of $p = s + 1$, the energy index of the
corresponding synchrotron emission is $\alpha = s/2$ ($L_\nu
\propto \nu^{-\alpha}$).  Depending on the strength of the magnetic
field, the energy index of the synchrotron spectrum just below
$\e_\peak$ will be either $\alpha = 1/2$ or $-1/3$.  If the field
strength is well below equipartition, i.e., $\xi_B \ll 1$, then the
cooling time will be very long, the electrons at the low energy
cut-off will have little time to cool, the flat portion of the
spectrum will be very narrow or non-existent, and the low energy
spectrum will be that of a cut-off electron distribution: $j_{\rm syn}
\propto \e^{1/3}$ (Katz 1994; Rybicki \& Lightman 1979).  This latter
behavior is seen in the spectra of the bright bursts GRB~910503 and
GRB~910814, while a softer transition region with $j_{\rm syn} \propto
\e^{-1/2}$ between the harder $\alpha = -1/3$ section at lower
energies and the peak in $\nu F_\nu$ may be present in the spectrum of
GRB~910601 (Schaefer et al.\ 1998).  This may indicate that the
synchrotron cooling of this latter burst is relatively more efficient
than that of the other two bursts.

More quantitatively, the location of the spectral peak and the shape
of the spectrum above and below $\e_\peak$ place strong limits on
the equipartition parameter $\xi_B$, the initial bulk Lorentz factor
$\G_0$ and the injection spectral index $s$.  For $\dt > \dt_{,0}$,
the lower cut-off electron Lorentz factor is
\begin{equation}
\g_\Min = \g_0\left[1 + \g_0\frac{a_0 t_0}{\zeta-1}
         \left(1 - \pfrac{r}{r_d}^{1-\zeta}\right)\right]^{-1},
\label{gmin_of_r}
\end{equation}
where $a_0 = (16/3) c \sigma_T (m_p/m_e) n_\ext \xi_B \G_0^2$ and $\g_0
= \eta (m_p/m_e)\G_0$.  In deriving this expression, we assume that
the particle injection is only significant once the blast-wave reaches
the deceleration phase $r > r_d$.  For a burst spectrum in which
electron cooling is important at all electron energies, which occurs
if $t_\co \simeq 2 \G_0 \dt < (\g/\gdot)_\Min$, the minimum electron
Lorentz factor obeys
\begin{eqnarray}
\g_\Min \le \g_{i,\Min} &=& \eta \frac{m_p}{m_e}\G \\
                        &=& \g_0\pfrac{r}{r_d}^{-\zeta}.
\label{gmin_i_of_r}
\end{eqnarray}
Writing $r/r_d = 1 + (r - r_d)/r_d$ and expanding
equations~\ref{gmin_of_r} \&~\ref{gmin_i_of_r} to first order, we
obtain $a_0 \ga \zeta/\g_0 t_0$.  In terms of the relevant model
parameters,
\begin{equation}
\eta \xi_B \G_0^{4/3} \ga 0.09\, E_{52}^{-1/3} n_1^{-2/3} \zeta,
\end{equation}
where $E_{52} = E/10^{52}\,$erg and $n_1 = n_\ext/1\,$cm$^{-3}$.
Using this expression in conjunction with equation~\ref{peak_energy},
we find
\begin{eqnarray}
\G_0 &\la & 2.6 \times 10^2\, E_{52}^{1/20} n_1^{-1/20} \zeta^{-3/20}
             \eta^{-9/20} \e_\peak^{3/10}
\label{Gamma_limit}\\
\xi_B &\ga & 5.5 \times 10^{-5}\, E_{52}^{-2/5} n_1^{-3/5} \zeta^{6/5}
             \eta^{-2/5} \e_\peak^{-2/5}.
\label{xi_B_limit}
\end{eqnarray}
Thus, for burst spectra peaking at $\sim 1$~MeV and for which electron
cooling is important, we obtain an upper limit for the initial bulk
Lorentz factor of $\sim 300$.  Remarkably, this upper limit depends
only very weakly on the total burst energy and the density of the
ambient medium.  For bulk Lorentz factors less than the above limit,
the equipartition fields will be sufficiently strong so that the
energy spectral index spectral index just below $\e_\peak$ will
be $\alpha = 1/2$ and there will be a break at even lower energies
below which $\alpha = -1/3$.

Bulk Lorentz factors larger than the above limit are possible, but in
these cases, the electrons will not be fully cooled.  If the injection
index is $s > 3$, then just above $\e_\peak$ the energy spectral index
will be that of the uncooled power-law electron distribution, $\alpha
= (s-1)/2$.  Furthermore, if the upper injection cut-off,
$\g_{i,\Max}$, is sufficiently high, an additional spectral break of
$\Delta\alpha = 1/2$ will exist at higher energies marking the
transition between the cooled and uncooled electrons (Dermer \& Chiang
1998).  Such a break would manifest itself as curvature in the high
energy tail above $\e_\peak$.  If the electron injection index is $s <
3$, then the peak in $\nu L_\nu$ may actually correspond to the
transition between cooled and uncooled power-law electron
distributions.  More likely, however, the peak in $\nu L_\nu$ will
correspond to the synchrotron emission from the electrons at the upper
cut-off (unless this upper cut-off energy is very high).  In this
case, there will be a spectral break between $\alpha = -1/3$ and
$\alpha = (s-1)/2$ below the peak in $\nu L_\nu$ and an exponential
cut-off above the peak (Rybicki \& Lightman 1979).  None of the bursts
for which good spectra are available show an exponential cut-off above
the $\nu L_\nu$ peak or a spectral index change of $\Delta\alpha =
1/2$ at the peak itself (Tavani 1996; Schaefer et al.\ 1998).  In
figure~\ref{burst_spectra}, we show burst spectra illustrating the
various spectral shapes we have described.

The evolution of the electron distributions will also be reflected in
the afterglow emission.  In figure~\ref{standard_calc}d, we show the
light curves at radio, optical, X-ray, gamma-ray, and TeV
(dot-dot-dot-dashed curve) energies.  As with our earlier calculation,
these energies are indicated by vertical lines in the
figure~\ref{standard_calc}b.  As the low energy electron cut-off
evolves downward due to cooling, the hard, low energy part of the
synchrotron spectrum will also move downward in energy.  The evolution
of the bulk Lorentz factor as the blast-wave decelerates contributes
to this effect as well.  At lower observer energies, the light curve
will peak later and the spectral index will get softer with time.
However, the precise shape of the light curve in any given band is
difficult to predict due to the uncertainties in the nature of the
electron cut-off and the evolution of the magnetic field.  The general
trend of hard-to-soft evolution should nevertheless hold, even if only
qualitatively.  Once the blast-wave has evolved so that the emission
of the cooled power-law electron distribution lies within the optical
band-pass, a measurement of the electron distribution power-law index
can be inferred from optical afterglow data.  For the afterglow of
GRB~970508, the optical energy spectral index is $\alpha \sim 0.7 \pm
0.3$ (Djorgovski et al.\ 1997), which is consistent with a cooled
power-law distribution with $s = 1.4$.  This value is quite different
from the injection indices implied by the prompt gamma-ray emission of
GRBs 910503, 910601 \& 910814 (Schaefer et al.\ 1998).  The energy
spectral indices for the high energy tail of the prompt burst emission
range from $\alpha = 1.3$--2.5, corresponding to injection indices of
$s = 2.6$--5 if these electrons are cooled and even larger values of
$s = 3.6$--6 if they are not cooled (Tavani 1996).

In addition to the evolution of the electron distribution injection
index, it is expected that the magnetic field will not stay at a
constant fraction of its equipartition value.  If there is some sort
of time scale, $\tau_{eq}$, for particle-field equipartition to
obtain, then one might expect that $\xi_B \approx \tau_{eq}/t_0$ where
$t_0 = r_d/c\G_0$ is the dynamical time scale of the blast-wave
expansion in the co-moving frame.  From equation~\ref{peak_energy},
bursts which peak at 1~MeV have weak magnetic fields with $\xi_B
\approx 10^{-4}$ for $\G_0 = 300$.  This implies a rather long 
equipartition time scale of $\tau_{eq} \approx 2\times 10^7$~s.  Even
for a non-radiative burst, this time scale is reached only after the
burst has almost completely decelerated.  Most likely, particle-field
equilibration is a highly non-linear process and thus cannot be
parametrized as a simple linear function of time.  The lack of a
proper theory describing such processes poses a major uncertainty in
using the characteristics of the prompt gamma-ray burst to set model
parameters for extrapolation of the behavior of the blast-wave to
describe the emission at much later times.

Nonetheless, for purposes of illustration, we present a calculation
which largely matches the observed properties of typical prompt GRB
emission.  In as much as the spectral slopes for GRBs below the $\nu
F_\nu$ peak are consistent with $\alpha = -1/3$ (Schaefer et al.\
1998), the limiting values of $\G_0 = 260$ and $\xi_B = 5.5 \times
10^{-5}$ apply.  These values also insure that for the prompt burst
spectrum we have $\e_{\rm peak}\sim 1$~MeV.  We have also set $E_0 =
10^{52}$~ergs, and used a compromise value for the injection index of
$s = 3$.  In figure~\ref{best_bet}, we show the results of a
calculation for those parameters.  The blast-wave is almost
non-radiative, with $\zeta \simeq 1.6$.  Despite the aforementioned
uncertainties in the model, this calculation displays many of the
properties which have been seen in actual bursts and their putative
afterglows.  In comparison with the calculation shown in
fig.~\ref{standard_calc}, it demonstrates the variety of afterglow
behaviors which can be obtained.

In figure~\ref{best_bet}b, the gamma-ray emission in the prompt burst
extends up to 1~GeV, well into the EGRET energy range (cf.\ Dingus
1995), the initial rises in the X-ray and optical light curves follow
that of the prompt burst itself by $\sim 10^3$~s and $\sim 5 \times
10^4$~s, respectively.  However, because of the steeper injection
index implied by the prompt burst, the flux in the X-ray band decays
as $\sim t^{-1.9}$, significantly faster than that measured for
GRB~970508 (Piro et al.\ 1998).  The optical light curve decays
somewhat slower as $\sim t^{-1.3}$ since the optical emission
corresponds to the flatter part of the spectrum near the peak in $\nu
L_\nu$.  The prompt burst spectrum evolves from hard to soft, and its
time scale is $\sim 10$~seconds.  The radio light curve is rising as
$\sim t^{0.3}$ and the radio spectral index is $\alpha = -1/3$.
Furthermore, this model makes the prediction of TeV emission due to
SSC processes which should peak at a time comparable to the peak in
the X-rays.

\section{Conclusions}

In this paper, we have attempted a more realistic calculation of the
dynamics and synchrotron and synchrotron self-Compton emission for the
blast-wave model of gamma-ray bursts.  By matching the detailed
characteristics of burst spectra, we have found relations
(eqs.~\ref{Gamma_limit} \&~\ref{xi_B_limit}) which place constraints
on magnetic field strengths and bulk Lorentz factors.  If these
relations are to be believed, then burst data can have a significant
impact on models of magnetic field generation in turbulent plasmas.

The apparent deficiencies of this calculation point towards areas of
further research.  In particular, the detailed temporal structure of
individual bursts is not explicitly dealt with in this model.  In the
context of external shocks, it may be due to inhomogeneities in the
external medium, or fluctuations in the electron injection and/or
magnetic field equipartition parameters (Panaitescu \& \Meszaros\
1998a).  For $s > 3$, the generalized expression for the luminosity for
energies $\e \ge \e_\peak$ (eq.~\ref{peak_luminosity}) is
\begin{equation}
\e^2 \frac{dN}{d\e dt} = 2\pi m_p c^2 \xi_e r^2 n_\ext(r) \G^2(r) 
                (\e/\e_\peak)^\lambda
\end{equation}
where $\lambda = (2-s)/2$ applies for relatively strong magnetic
fields $\xi_B \ga 10^{-4}$ when the electrons just above the break are
efficiently cooled and $\lambda = (3-s)/2$ applies for relatively weak
fields and uncooled electrons.  From this expression we see that any
burst light curve substructure must be due to variations in $\xi_e$
and $n_\ext$, and indirectly, due to variations in $\xi_B$ through
$\e_\peak$ (eq.~\ref{peak_energy}).

Our treatment of the dynamics also ignores the structure of the shock
region itself, and our approach essentially only considers the
emission from the forward shock and neglects the reverse shock.
Panaitescu \& \Meszaros\ (1998a) have performed calculations similar
to our own, but from a hydrodynamical perspective, and found that the
reverse shock only makes a significant contribution to the emission at
optical and UV energies.  Therefore, neglecting the reverse shock
should not affect our results for the gamma-ray emission, but it could
have a significant impact on the optical and radio afterglow emission.
We also neglect the thickness, $\Delta r$, of the shock shell in
integrating the emission for a given observer time $\dt$.  This should
not be important at early times when $\Delta r = r_0/\G_0^2$ (in the
lab frame), but it could affect the afterglow emission at late times.

It is unlikely that the blast-wave itself is spherical.  If the
initial fireball is created by the coalescence of two compact objects,
then the orbital plane defines a natural axis of symmetry along which
the blast-wave will propagate (\Meszaros\ \& Rees 1992b).  This sort
of asymmetry could be accounted for in our model by a non-unity
collimation factor, $f_b$.  Furthermore, if the observer line-of-sight
does not lie within the opening angle of the blast-wave cone, then
other effects due to relativistic beaming and the gradual deceleration
of the shock front would be introduced.  In this respect, highly
anisotropic blast-waves would share properties with relativistic jets
in blazars.

This analogy can be take even further by noting the similarity of the
burst spectra we derive compared to that of gamma-ray blazars.  Like
our model spectra, the spectral energy distributions (SEDs) of these
objects tend to have two peaks, one in the UV--X-ray range and one at
gamma-ray energies.  If the lower peak in blazar SEDs is due to
synchrotron emission and corresponds to the $\sim 1$~MeV peak in
gamma-ray bursts, we can apply a similar analysis as we have discussed
above to derive bulk Lorentz factors and equipartition parameters for
blazars.  In particular, the recent ASCA observations of Mrk~421
(Takahashi et al.\ 1996) provide sufficient information to get actual
values rather than simply upper or lower limits.  Using the light
curves of Mrk~421 measured in different X-ray energy bands, Takahashi
et al.\ (1996) performed a cross-correlation analysis and found that
the longer relative time lags of the lower energy data versus the
higher energy data are consistent with synchrotron cooling of the
underlying electron distribution.  Several authors have noted this
effect and have calculated this temporal dependence for the cases of
bursts and blazars (e.g., Tashiro et al.\ 1995; Tavani 1996; Dermer
1998).  Takahashi et al.\ use the TeV variability time scale (Kerrick
et al.\ 1995) to estimate a Doppler factor and find $\D = 5$
(cf.\ Takahara 1994).  Using this estimate and their time lag
measurements, they derive a magnetic field of $B = 0.2$~G. From
non-simultaneous data (Shrader \& Wehrle 1997), the synchrotron
portion of the SED of Mrk~421 peaks at about $\sim 130~$eV. Using
\begin{equation}
\e_\peak = \frac{B}{B_{\rm crit}} \g^2 {\cal D}
\end{equation}
(cf.\ eq.~\ref{peak_energy}), and $\g = (m_p/m_e)\G$, we find $\G
\approx 60$ and an observer angle $\theta \approx 5^\circ$.
We also obtain an equipartition field strength of $B_{eq} \approx 10
n_1^{1/2}$~G implying an equipartition parameter of $\xi_B \sim
10^{-2}$.  Although the above value for the bulk Lorentz factor is
substantially larger than the mean value of $\langle \G \rangle \sim
10$ found by applying the beaming model to a sample of radio-loud
objects (Urry \& Padovani 1995), its large value may indicate the
special nature of gamma-ray loud blazars which are characterized not
only by small observing angles but also by larger than typical bulk
Lorentz factors.  Despite the crudeness of this calculation, it
illustrates the potential applicability of this sort of analysis to
blazars as well as bursts.

\acknowledgements
We thank Jeff Skibo for many useful discussions, and we thank Markus
B\"ottcher for providing us with his computer code for calculating the
synchrotron emissivity.  This work was performed while J.C. held a
National Research Council-NRL Research Associateship and was supported
by the Office of Naval Research and the {\em Compton Gamma-Ray
Observatory} Guest Investigator program.

\begin{figure}[p]
%\epsfysize=6in
%\centerline{\epsfbox{test_case.epsf}}
\caption{Comparison with the analytic result of Dermer \&~Chiang 1998.  
The parameters are $\G_0 = 300$, $B = 1$~G, $n_\ext = 1$~cm$^{-3}$, $s
= 2$.  {\em Upper panel:} The spectra, from top to bottom, are at
observer times $\dt = 1, 10, 10^2, \ldots, 10^6$~seconds, with the
present calculation plotted as the solid curves and the analytic
estimate as the dashed curves.  {\em Lower panel:} Light curves at
radio (2.4~GHz, dotted), optical (1~eV, solid), X-ray (1~keV, dashed)
and gamma-ray (1~MeV, dot-dashed) energies.  The results from the
analytic estimate are over-plotted as symbols.  Light curves for each
successive energy band are displaced vertically by 2 units for
clarity.}
\label{test_case}
\end{figure}

\begin{figure}[p]
%\epsfysize=6in
%\centerline{\epsfbox{surface_brightness.epsf}}
\caption{{\em Upper panel:} Radio surface brightness versus perpendicular 
distance as a function of observer angle $\theta$ at constant time
delays, $\dt = 1, 10, 10^2, \ldots$~seconds (top to bottom).  {\em
Lower panel:} Apparent size versus time delay.  The data points are
from the above calculation and the solid curve is
equation~\protect{\ref{burst_size}} assuming $\zeta = 3$, i.e., a
radiative blast-wave.}
\label{surface_brightness}
\end{figure}

\begin{figure}[p]
%\epsfxsize=\hsize
%\centerline{\epsfbox{standard_calc.epsf}}
\caption{Numerical results from our blast-wave calculation using 
model parameters for a burst in equipartition, i.e., $\xi_B = 1$.
Other relevant parameters are $\xi_e = 0.5$, $\G_0 = 300$, $E_0 =
10^{52}$~ergs, $n_\ext = 1$~cm$^{-3}$, $s = 3$, and $\eta = 1$.  {\em
Panel~a:} The bulk Lorentz factor of the blast-wave shell as a
function of shell radius.  The vertical dotted lines indicate the
radii of the nearest surface of the shell at observer times $\dt =
1,10,10^2,\ldots$~s.  The power-law of the bulk Lorentz factor during
the deceleration phase is $\zeta \simeq 1.9$.  {\em Panel~b:}
Synchrotron and synchrotron self-Compton spectra at the above observer
times.  The spectrum corresponding to each time can be identified by
the energy, $\e_\peak$, of its synchrotron peak which decreases
monotonically with time.  As in figure~\protect{\ref{test_case}}, the
vertical lines denote various observer energies---the
dot-dot-dot-dashed line corresponds to 1~TeV.  {\em Panel~c:}
Co-moving electron distributions at the leading surface of the shell
at the above observer times.  Each distribution can be identified by
the high energy break which corresponds to the synchrotron peak and
which also decreases monotonically with time.  The dashed curves show
our analytic estimate (eq.~\protect{\ref{flat_spec}}) for these
distributions.  {\em Panel~d:} Light curves for radio, optical, X-ray,
gamma-ray and TeV energies.}
\label{standard_calc}
\end{figure}

\begin{figure}
%\epsfxsize=\hsize
%\centerline{\epsfbox{burst_spectra.epsf}}
\caption{Burst spectra near $\e_\peak$ for different values of 
the magnetic field equipartition parameter illustrating the effect of
different degrees of synchrotron cooling on the burst spectral shape.
In all three cases, the initial bulk Lorentz factor $\G_0 = 260$
(eq.~\protect{\ref{Gamma_limit}}).  For the solid, dashed and dotted
curves we us an electron injection index $s=4$.  {\em Solid curve:}
The spectrum obtained for the limiting value $\xi_B = 5.5 \times
10^{-5}$ (eq.~\protect{\ref{xi_B_limit}}) which marks the transition
between complete and incomplete electron cooling.  The energy spectral
indices ($L_\nu \propto \nu^{-\alpha}$) above and below $\e_\peak$ are
$\alpha = -1/3$ and $\alpha = s/2 = 2$ respectively.  {\em Dashed
curve:} Here $\xi_B = 1 \times 10^{-6}$, and we have $\alpha = (s-1)/2
= 3/2$ above $\e_\peak$.  {\em Dotted curve:} $\xi_B = 1 \times
10^{-2}$.  Above $\e_\peak$, $\alpha = s/2 = 2$, and below $\e_\peak$
there is an additional spectral break due to the cooled electrons
which have been injected at the lower cut-off energy
(eq.~\protect{\ref{flat_spec}})---just below $\e_\peak$, $\alpha =
1/2$ and below the additional break, $\alpha = -1/3$.  {\em Dot-dashed
curve:} $\xi_B = 10^{-5}$ and $s = 2$.  Just above the low energy
break we have $\alpha = (s-1)/2 = 1/2$, and the $\nu L_\nu$ peak
corresponds to emission from the electrons at the high energy
cut-off.}
\label{burst_spectra}
\end{figure}

\begin{figure}[p]
%\epsfxsize=\hsize
%\centerline{\epsfbox{best_bet.epsf}}
\caption{Burst calculation with $\G_0 = 260$ and 
$\xi_B = 5.5 \times 10^{-5}$.  All other parameters are the same as
for figure~\protect{\ref{standard_calc}}.  The thin curve in panel~a
is the initial estimate of the bulk Lorentz factor using
eq.~\protect{\ref{analytic_Gamma}}.  The thick curve is the solution
to which our iterative procedure converges.  The difference in these
curves shows the effect of the electrons not being radiative due to
the weak magnetic fields.}
\label{best_bet}
\end{figure}
\clearpage

\setcounter{figure}{0}
\begin{figure}[p]
\epsfysize=6in
%\centerline{\epsfbox{test_case.epsf}}
\centerline{\epsfbox{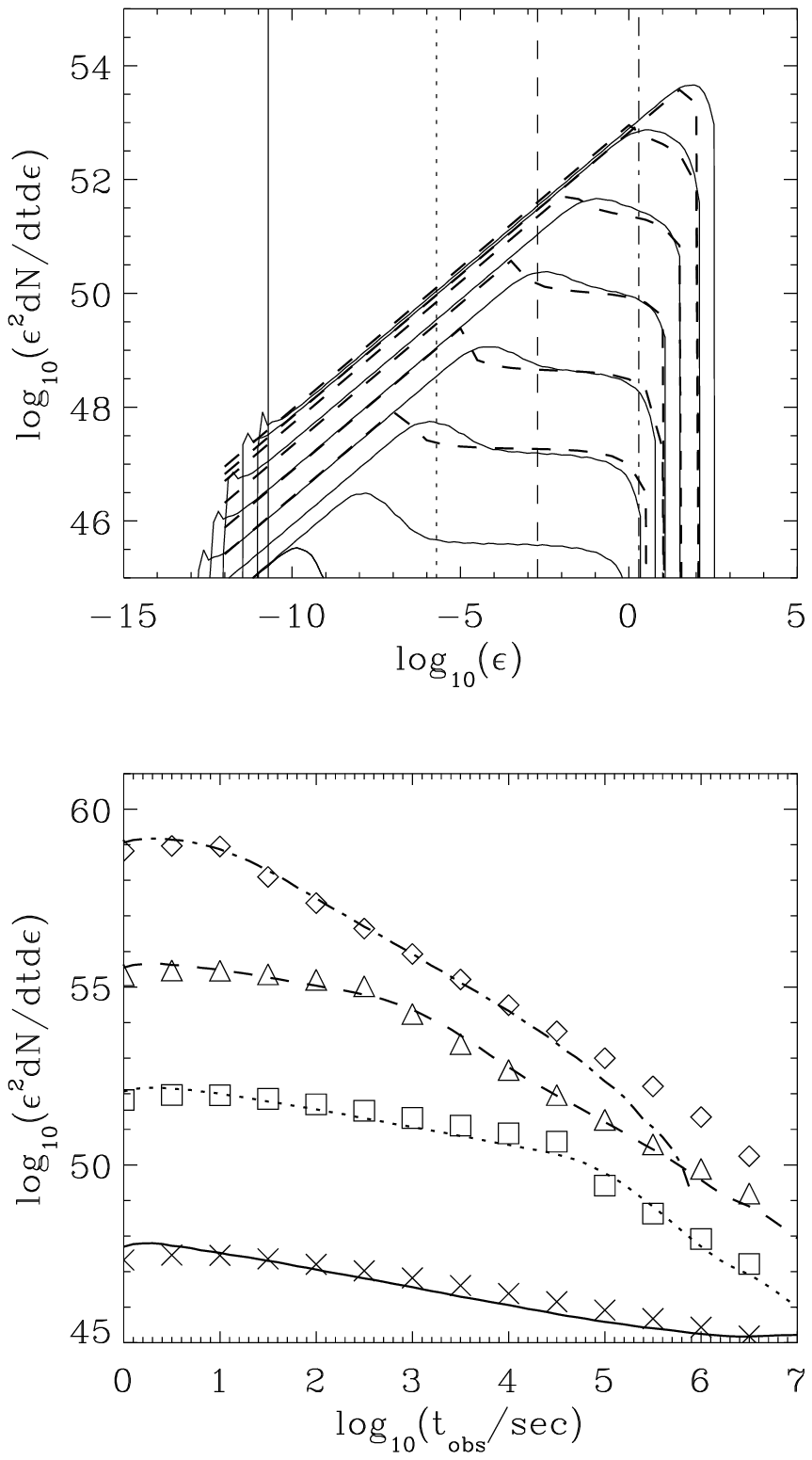}}
\caption{ }
\end{figure}

\begin{figure}[p]
\epsfysize=6in
%\centerline{\epsfbox{surface_brightness.epsf}}
\centerline{\epsfbox{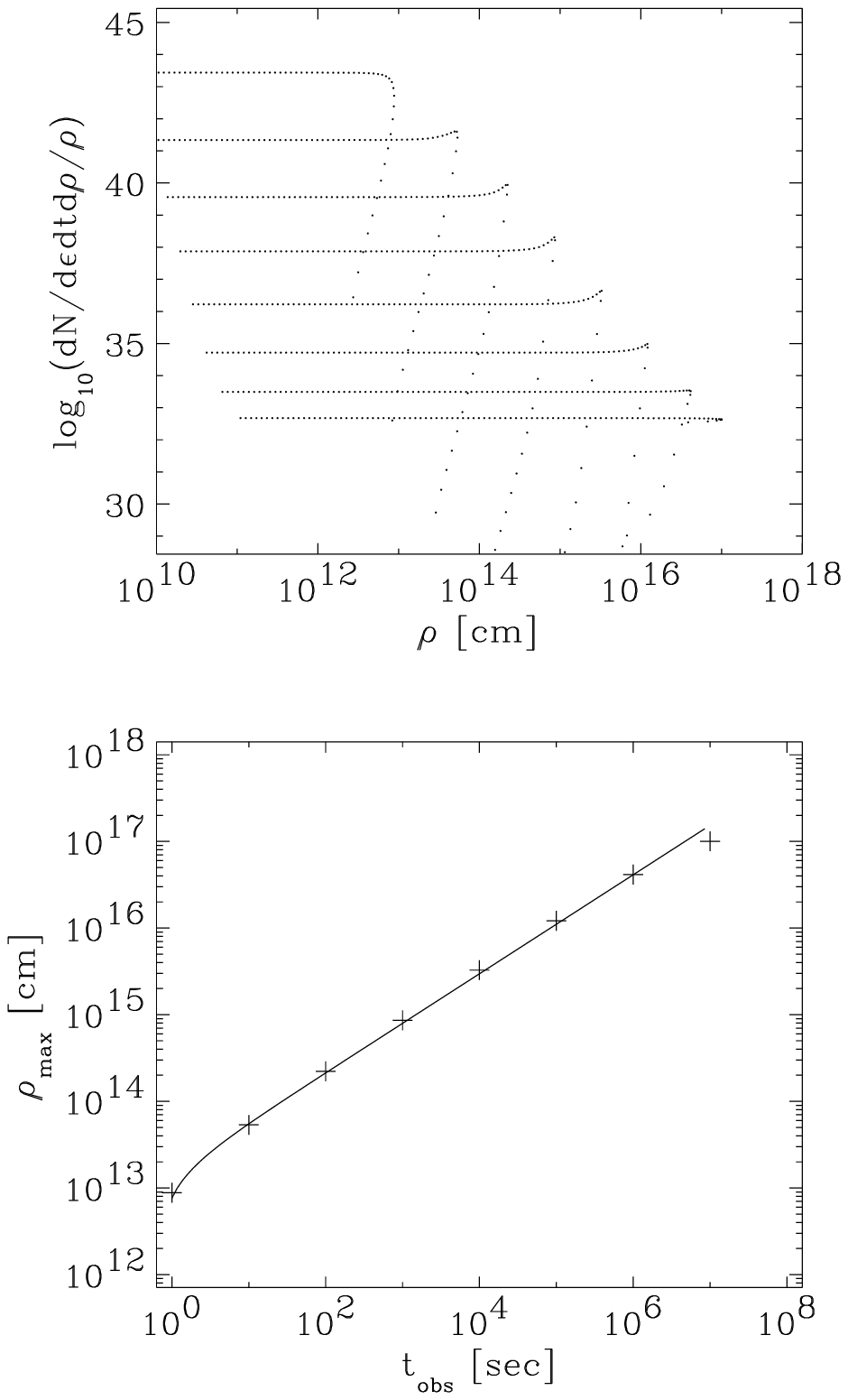}}
\caption{ }
\end{figure}

\begin{figure}[p]
\epsfxsize=\hsize
%\centerline{\epsfbox{standard_calc.epsf}}
\centerline{\epsfbox{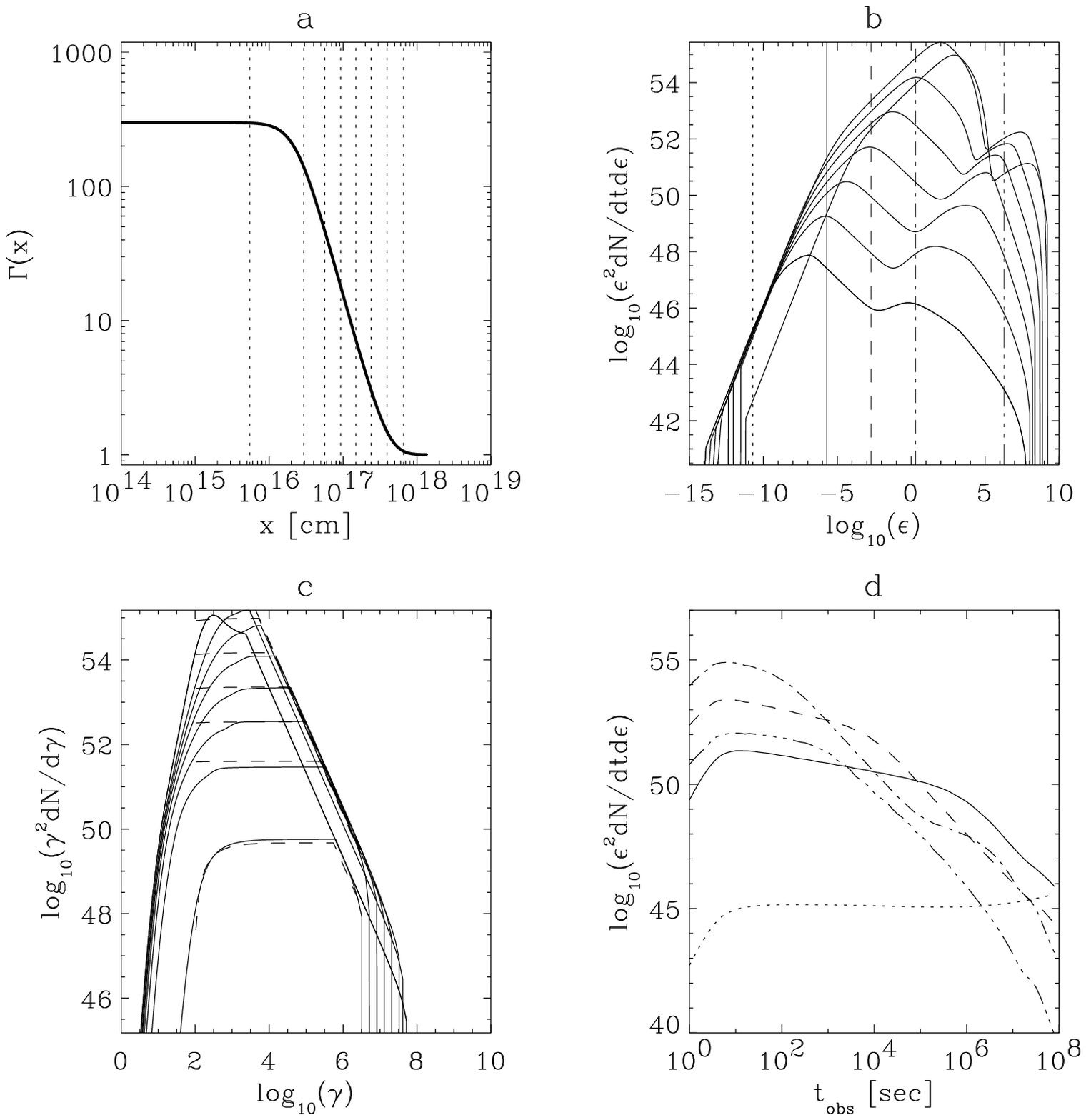}}
\caption{ }
\end{figure}

\begin{figure}[p]
\epsfxsize=\hsize
%\centerline{\epsfbox{burst_spectra.epsf}}
\centerline{\epsfbox{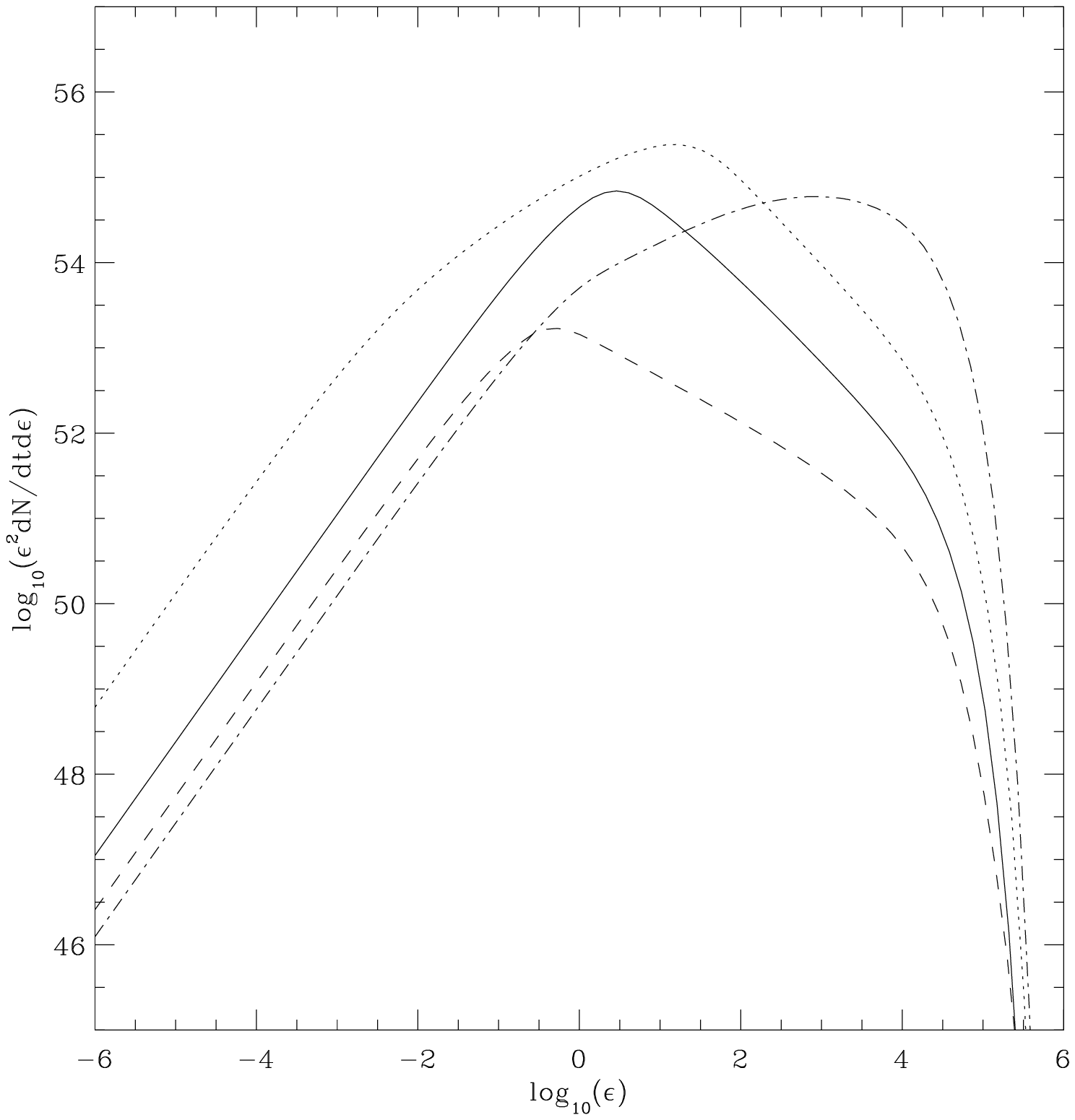}}
\caption{ }
\end{figure}

\begin{figure}[p]
\epsfxsize=\hsize
%\centerline{\epsfbox{best_bet.epsf}}
\centerline{\epsfbox{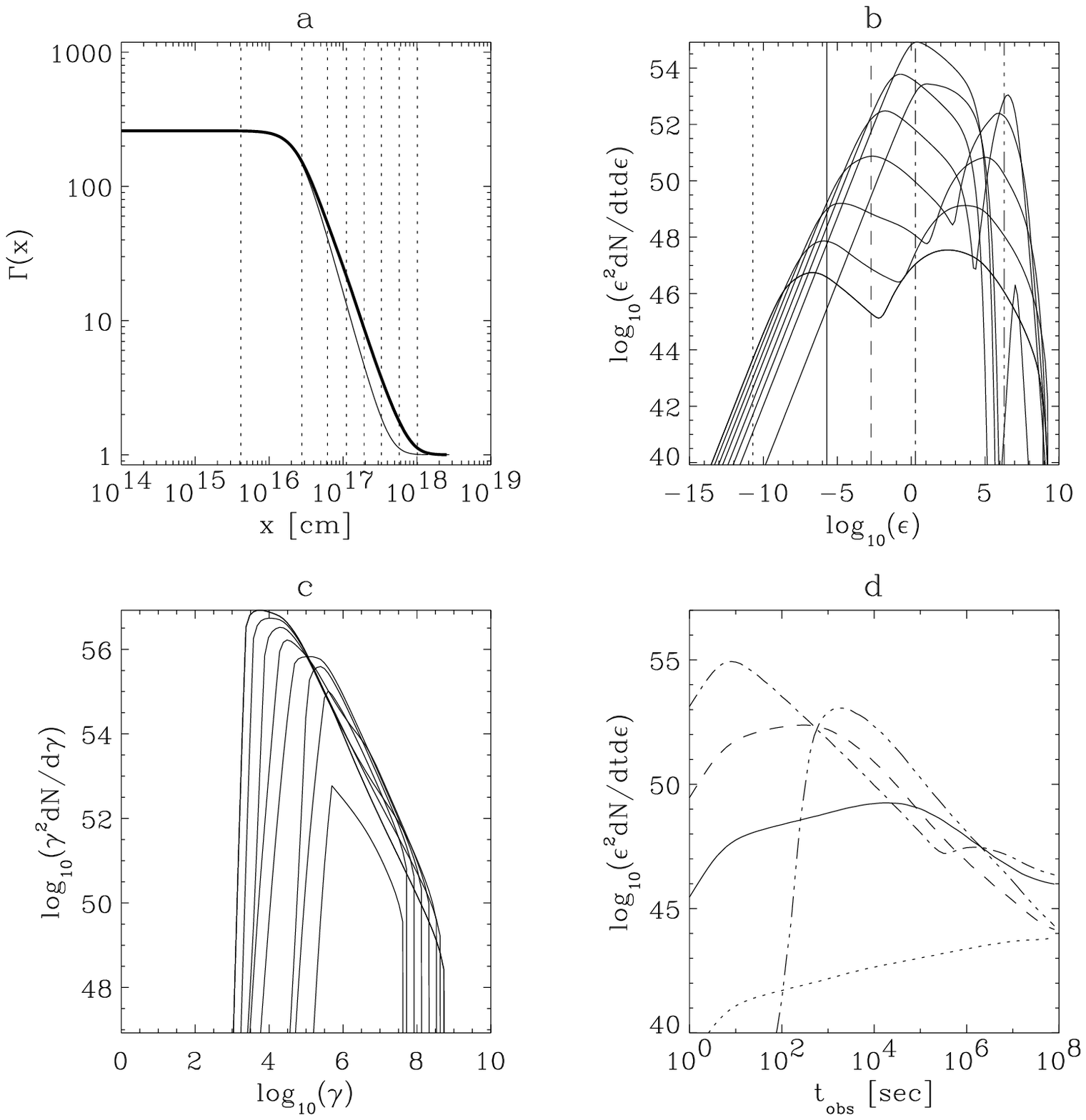}}
\caption{ }
\end{figure}

\end{document}